\documentclass[prl,reprint]{revtex4-1}
\usepackage{amsmath}
\usepackage{graphicx}
\usepackage{color}
\usepackage{soul}

\begin{document}

\title{Induced interactions in a superfluid Bose-Fermi mixture}
\author{J. J. Kinnunen}
\affiliation{COMP Centre of Excellence and Department of Applied Physics, Aalto University, FI-00076 Aalto, Finland}
\author{G. M. Bruun}
\affiliation{Department of Physics and Astronomy, Aarhus University, DK-8000 Aarhus C, Denmark}

\begin{abstract}
We analyse a Bose-Einstein condensate (BEC) mixed with a superfluid two-component Fermi gas in the whole BCS-BEC cross-over. 
Using a quasiparticle random phase approximation combined with Beliaev theory to describe the Fermi superfluid and the BEC respectively, we show that the 
single particle and collective excitations of the Fermi gas give rise to an induced interaction between the bosons, which varies strongly with momentum and frequency.
It diverges at the sound mode of the Fermi superfluid, resulting in a sharp avoided crossing feature and a corresponding sign change 
 of the interaction energy shift in the excitation spectrum of 
the BEC.
In addition, the excitation of quasiparticles in the Fermi superfluid leads to damping of the excitations in the BEC. 
Besides studying induced interactions themselves, these prominent effects can be used to systematically probe the strongly interacting Fermi gas. 
\end{abstract}
\maketitle


The interplay between induced interactions and superfluidity plays an important role in low temperature physics. 
In metals, the phonon mediated interaction between electrons leads to the formation of Cooper pairs~\cite{Schrieffer}, and induced electron-hole excitations significantly suppress the critical temperature of a BCS superconductor~\cite{Heiselberg2000,Gorkov}. 
A prominent theory for high temperature superconductivity is that 
it is caused by spin fluctuations leading to an attractive interaction~\cite{Scalapino}, and induced interactions are important for 
understanding the properties of liquid helium mixtures~\cite{BaymPethick}. 
The systems where induced interactions are significant often consist of fermionic and bosonic degrees of freedom. 
In cold atom gases, Bose-Fermi mixtures have been realised experimentally for sympathetic cooling~\cite{Truscott,Schreck,Roati}, 
molecule formation~\cite{Wu,Park,Heo,Cumby}, and for studying few-body physics~\cite{Tung}. 
The theoretical studies have focused on mixtures where the Fermi gas is in the normal state~\cite{Viverit2000,Bijlsma2000,Yu,Ludwig,Fratini,Bertaina,Sogo,Pixley,Santamore2008,Albus2002}. 
Recently, an  experimental breakthrough was reported with the realisation of a mixture of superfluid $^7$Li and $^6$Li gases~\cite{Barbut}. 
This opens up the exciting possibility to experimentally study for the first time the role of induced interactions in a Bose-Fermi mixture, where both components are superfluid.

Here, we study a BEC mixed with a two-component superfluid Fermi gas in the whole BCS-BEC crossover at zero temperature.
Using a  quasiparticle random-phase approximation (QRPA) to describe the excitations in the Fermi gas, combined with Beliaev theory for the bosons, we show 
how the fermions give rise to an induced frequency/momentum dependent Bose-Bose interaction, which diverges at the sound mode of the  Fermi gas. 
This results in two qualitatively new effects. 
First, the dispersion relation of the bosons in the BEC is significantly changed  
at frequencies/momenta close to the sound mode of the Fermi gas. 
Second, bosonic excitations are damped due to dissipation, as they can excite quasiparticles in the superfluid Fermi gas~\cite{Zheng2014a}. 
These effects can be used to systematically probe the single particle and collective properties of the strongly correlated Fermi gas.

We consider a gas of bosons with  mass $m_{\rm B}$ mixed with a two-component ($\sigma=\uparrow,\downarrow$) gas of fermions with mass $m_{\rm F}$. 
The populations of the two fermionic  states are taken to be the same. 
The Hamiltonian of the Bose-Fermi mixture is $H=H_{\rm B}+H_{\rm F}+H_{\rm BF}$, where 
\begin{equation}
H_{\rm B}=\sum_{{\mathbf k}}\epsilon_ka^\dagger_{{\mathbf k}}a_{{\mathbf k}}+\frac 1 {2{\mathcal V}} \sum_{{\mathbf k},{\mathbf k}',{\mathbf q}}V_{\rm B}(q)a^\dagger_{{\mathbf k}+{\mathbf q}}a_{{\mathbf k}'-\mathbf{q}}^\dagger a_{{\mathbf k}'} a_{{\mathbf k}}
\end{equation}
is the Bose hamiltonian with $\epsilon_k=k^2/2m_{\rm B}$, 
\begin{equation}
H_{\rm F}=\sum_{{\mathbf k}\sigma}\frac{k^2}{2m_{\rm F}}c^\dagger_{{\mathbf k}\sigma}c_{{\mathbf k}\sigma}+\frac 1 {{\mathcal V}} \sum_{{\mathbf k},{\mathbf k}',{\mathbf q}}V_{\rm F}(q)
c^\dagger_{{\mathbf k}+{\mathbf q}\uparrow}c_{{\mathbf k}'-\mathbf{q}\downarrow}^\dagger c_{{\mathbf k}'\downarrow} c_{{\mathbf k}\uparrow}
\end{equation}
is the Fermi Hamiltonian, and 
\begin{equation}
H_{\rm BF}=\frac 1 {{\mathcal V}} \sum_{{\mathbf k},{\mathbf k}',{\mathbf q}\sigma}V_{\rm BF}(q)c_{{\mathbf k}+\mathbf{q}\sigma}^\dagger c_{{\mathbf k}\sigma}a^\dagger_{{\mathbf k}'-{\mathbf q}} a_{{\mathbf k}'}
\label{HBF}
\end{equation}
is the Bose-Fermi interaction. 
The operators $a_{\mathbf k}$ ($c_{{\mathbf k}\sigma}$) remove a boson (spin $\sigma$ fermion) with momentum ${\mathbf k}$,  ${\mathcal V}$ is the volume of the system, and 
 we work in units where $\hbar=k_{\rm B}=1$.
In the following, we replace the interactions with the corresponding low energy scattering matrices: 
$V_{\rm B}(q)\rightarrow {\mathcal T}_{\rm B}=4\pi a_{\rm B}/m_{\rm B}$,  $V_{\rm F}(q)\rightarrow {\mathcal T}_{\rm F}=4\pi a_{\rm F}/m_{\rm F}$, and 
 $V_{\rm BF}(q)\rightarrow {\mathcal T}_{\rm BF}=2\pi a_{\rm BF}/m_{\rm r}$, where $a_{\rm B}$, $a_{\rm BF}$, and $a_{\rm F}$ is the Bose-Bose, Bose-Fermi, and Fermi-Fermi scattering length respectively, and $m_{\rm r}=m_{\rm B}m_{\rm F}/(m_{\rm B}+m_{\rm F})$ is the reduced mass. As usual, this corresponds to summing all ladder diagrams in a vacuum.   

\begin{figure}
\includegraphics[width=0.98\columnwidth]{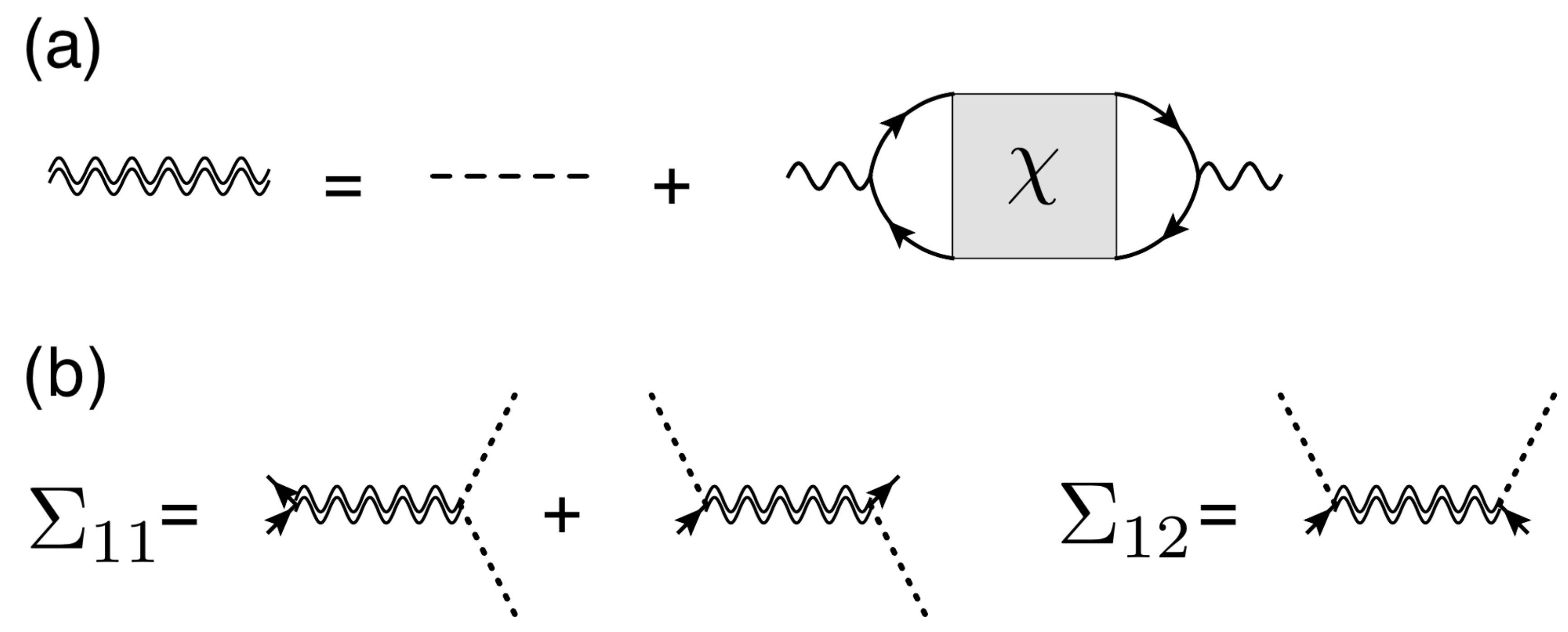}
\caption{(Color online) (a) The effective interaction  $V(q,\omega)$ (double wavy line) between the bosons. The dashed line is the bare 
Bose-Bose interaction ${\mathcal T}_{\rm B}$, single wavy lines are the Bose-Fermi interaction ${\mathcal T}_{\rm BF}$, and the solid lines are the Fermi Green's function. 
(b) The Bose self-energies $\Sigma_{11}(q,\omega)$ and 
$\Sigma_{12}(q,\omega)$. The dotted lines are excitations in and out of the BEC.  } 
\label{UefFig}
\end{figure}

The presence of the Fermi gas induces an effective interaction between the bosons, since one boson tends to attract/repel fermions giving rise to a local change in the fermion density, which is felt by the second boson.  Combined with the direct Bose-Bose interaction, this results in total interaction
\begin{equation}
   V(q,\omega) = {\mathcal T}_{\rm B} + {\mathcal T}_{\rm BF}^2\chi(q,\omega).
   \label{Vtot}
\end{equation}
Here $\chi(q,\omega)$ is the density-density response function for the fermions with momentum ${\mathbf q}$, and frequency $\omega$.
The corresponding Feynman diagram for $V(q,\omega)$ is given in Fig.~\ref{UefFig}~(a).
The momentum dependence reflects  the long-range of the interaction, as density perturbations propagate in the Fermi gas. 
Similarly, the frequency dependence of the interaction is due to the fact that it is not instantaneous since these perturbations have a finite speed.

In the weak coupling BCS limit, $k_{\rm F}a_{\rm F}\rightarrow 0_-$ with $k_{\rm F}$ the Fermi momentum of the Fermi gas, the density-density response function 
is given by
\begin{equation}
  \chi(q,\omega) =\left(\frac{v_{\rm F}}{c_{\rm s}}\right)^2\frac{\mathcal{N}(\epsilon_{\rm F})}{3[(\frac{\omega}{c_{\rm s}q})^2-1]}
\label{eq:weaksfresponseBCS}
\end{equation}
for frequency/momenta close to the Anderson-Bogoliubov sound mode $\omega=c_{\rm s}q$. 
The velocity is $c_{\rm s}=v_\mathrm{F}\sqrt{1 + 2k_\mathrm{F} a_\mathrm{F}/\pi}/\sqrt3$~\cite{Anderson,Bogoliubov,Minguzzi2001},  
the density of states at the Fermi level $\epsilon_{\rm F}=k_{\rm F}^2/2m_{\rm F}$ is $\mathcal{N}(\epsilon_{\rm F})=m_{\rm F}k_{\rm F}/\pi^2$, and $v_{\rm F}=k_{\rm F}/m$.
In the BEC regime, $k_{\rm F}a_{\rm F}\rightarrow 0_+$, the Fermi gas becomes a BEC consisting of diatomic molecules (dimers) 
with mass $2m_{\rm F}$ and density $n_{\rm F}/2$, where $n_{\rm F}=k_{\rm F}^3/3\pi^2$ is the total density of the fermions. The density-density response function
is then from Bogoliubov theory given by~\cite{NozieresPines}   
\begin{equation}
  \chi(q,\omega) = \frac{n_{\rm F} q^2}{4m_{\rm F}(\omega^2 - \omega_q^2)}\simeq\left(\frac{v_{\rm F}}{c_{\rm s}}\right)^2\frac{\mathcal{N}(\epsilon_{\rm F})}{12[(\frac{\omega}{c_{\rm s}q})^2-1]}.
\label{eq:weaksfresponseBEC}
\end{equation}
Here, $\omega_q^2=q^2(4m_{\rm F})^{-1}(q^2/4m_{\rm F}+0.6{\mathcal T}_{\rm F}n_{\rm F}/2)$ is the Bogoliubov spectrum of the dimer BEC, where we have used that  the scattering length between the dimers is $0.6a_{\rm F}$ in the BEC limit~\cite{Petrov}.
The second equality in~\eqref{eq:weaksfresponseBEC} follows from the fact that $\omega_q\simeq c_{\rm s}q$
for small momenta, where $c_{\rm s}=\sqrt{0.6a_{\rm F}n_{\rm F}\pi/2m_{\rm F}^2}$ is the Bogoliubov sound speed. 

In general, the density-density correlation function of the Fermi gas has a pole at $\omega=c_{\rm s}q$ in the whole BCS-BEC crossover, where $c_{\rm s}$ is the velocity of sound for a given scattering length $-\infty<a_{\rm F}<\infty$. 
It follows from~\eqref{Vtot} that the induced interaction between the bosons has the same pole structure: it is attractive for $\omega\le c_{\rm s}q$, 
repulsive for $\omega\ge c_{\rm s}q$,  
and it \emph{diverges} when $\omega=c_{\rm s}q$. 
In addition, it has a non-zero imaginary part for frequency/momenta inside the quasiparticle continuum of the Fermi gas. 
It also follows  from~\eqref{Vtot}-\eqref{eq:weaksfresponseBEC} that the strength $\kappa$ of the induced interaction scales as 
\begin{equation}
\kappa={\mathcal T}_{\rm BF}^2\mathcal{N}(\epsilon_{\rm F})\frac{v_{\rm F}^2}{c_{\rm s}^2}
\label{Couplingstrength}
\end{equation}
 which should be compared with the strength $ {\mathcal T}_{\rm B} $ of the direct Bose-Bose interaction.

We now examine the effects of the induced interaction on the excitation spectrum of the Bose gas.
To this end, we need to calculate the density-density response function of the Fermi gas in  the whole BCS-BEC regime. 
The density response function $\chi(1,2)$ is defined as a measure for how much the density of the Fermi gas changes at point (and time) $1$ 
when a potential perturbation $\delta \phi $ is applied at point $2$:
\begin{equation}
  \chi(1,2) = -\frac{\delta \langle n(1) \rangle}{\delta \phi(2)}.
\end{equation}
We apply a QRPA for calculating the Fourier transform of $\chi(1,2)$  in the superfluid state~\cite{CoteGriffin,Leggett,Engelbrecht,Bruun2001,Minguzzi2001}.
This is the simplest microscopic scheme which recovers the Anderson-Bogoliubov mode in the BCS regime, and the Bogoliubov mode in the BEC regime. 
It yields a response function of the form  
\begin{equation}
 \chi(q,\omega) = \frac{\chi_0(q,\omega)}{1-\mathcal{T}_{\rm F} L(q,\omega)},
 \label{chiBCS}
\end{equation}
where  $\chi_0(q,\omega)$ is a four-dimensional vector giving response due to quasiparticle excitations in the superfluid, 
and $L$ is a $4\times4$ matrix describing the  couplings of the densities and the order parameter field.
The collective modes manifest themselves as poles of the density response $\chi({ q},\omega)$, i.e. as the zeroes of the determinant
\begin{equation}
  \mathrm{det} \left[1 - \mathcal{T}_{\rm F} L({q},\omega) \right] = 0.
\end{equation}

The input parameters needed for the QRPA  are the chemical potential $\mu$ and the pairing gap $\Delta$ of the 
Fermi superfluid, which are obtained self-consistently from BCS theory. 
We have for convergence  added a small imaginary part $i\eta=i10^{-3}\,\epsilon_\mathrm{F}$ to the frequencies, and checked that the final numerical results do not depend on $\eta$, as long as $\eta\ll\epsilon_{\rm F}$. The details of this QRPA calculation can be found for example in Refs.~\cite{Minguzzi2001} and~\cite{CoteGriffin}.

\begin{figure}
\centering
\includegraphics[width=\columnwidth]{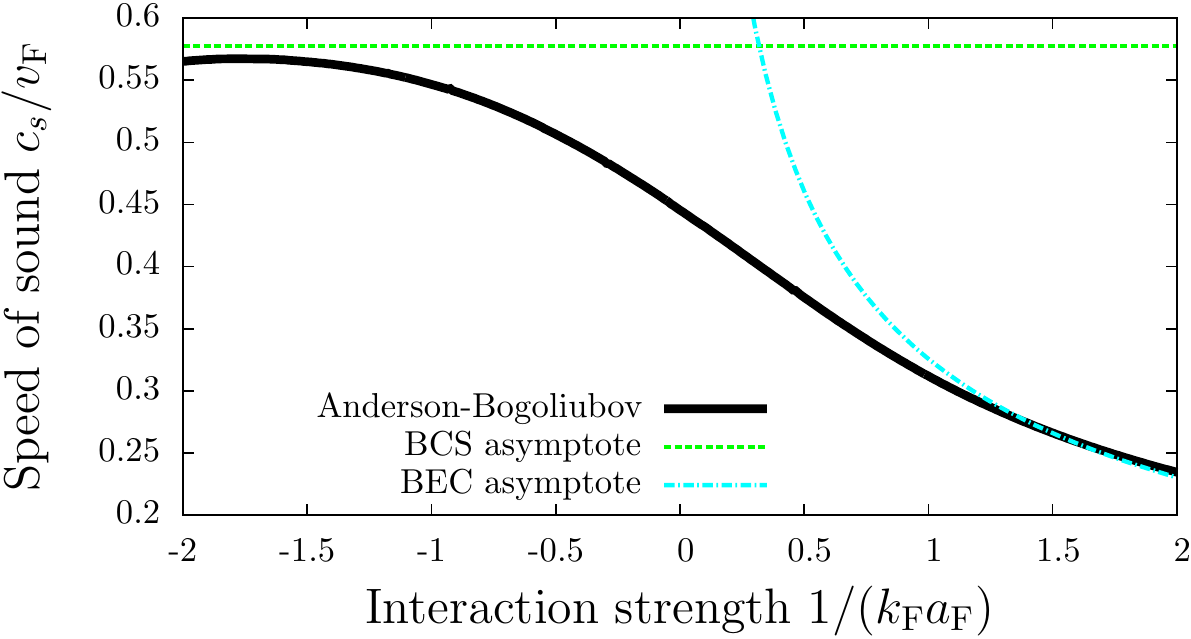}
\caption{(Color online) Speed of sound $c_s$ in the two-component Fermi superfluid as calculated from the pole of the density response function $\chi({\bf q},\omega)$. It
 approaches  $c_s=v_{\rm F}/\sqrt{3}$ in the BCS limit (the constant green dotted line), and the 
  Bogoliubov result $c_s = v_{\rm F}\sqrt{k_{\rm F}a_{\rm F}/3\pi}$ in the BEC limit (blue dash-dotted curve).}
\label{fig:speedofsound}
\end{figure}
Figure~\ref{fig:speedofsound} shows the speed of the Anderson-Bogoliubov mode as a function of $1/k_Fa$, determined 
by finding the frequency $\omega$ at which the imaginary part of $\chi(q,\omega)$  is maximal for a given momentum $q$.
The value of the  momentum $q$ needs to be chosen small enough so that it probes the linear part of the collective mode branch.
The speed of sound is then the slope $c_s = \omega/q$. 
In the BCS limit, the speed of sound approaches the weakly interacting limit $v_{\rm F}/\sqrt{3}$. 
The numerically calculated speed of sound deviates slightly from this in the very weakly interacting regime, due to the difficulty of determining the slope when the pairing gap is very small.
Our numerics reproduce to an excellent accuracy the speed of sound results  in Ref.~\cite{Combescot2006}.
Note that this theory is of course not quantitatively correct in the whole BCS-BEC crossover. 
For instance, the speed of sound approaches $c_s = v_{\rm F}\sqrt{k_{\rm F}a_{\rm F}/3\pi}$ in the BEC limit,  see Fig.~\ref{fig:speedofsound}. This corresponds to a 
 molecular BEC with a scattering length $2a_{\rm F}$, instead of the correct value $0.6a_{\rm F}$.  
We emphasize however, that the effects discussed below are completely general and do not depend 
on which approximate theory we apply to describe the strongly correlated system.

\begin{figure}
\includegraphics[width=\columnwidth]{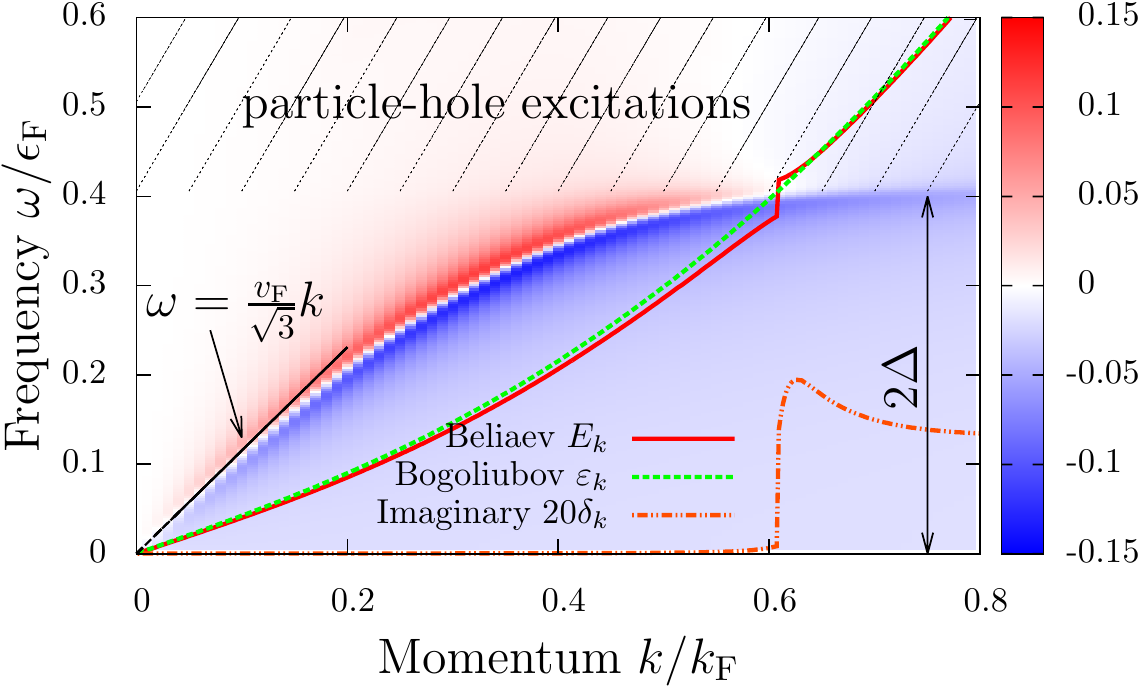}
\caption{(Color online) The blue and red regions show $\mathrm{Re}\,\chi(q,\omega)$ for $k_{\rm F}a_{\rm F}=-1$. 
The black solid line is the weak coupling Anderson-Bogoliubov mode, and the quasiparticle continuum for  $\omega > 2\Delta$ is indicated by a dashed region. 
The green dashed line is the Bogoliubov  spectrum  $\varepsilon_k$ of the atomic BEC, and the red solid line is the Beliaev spectrum $E_k$ for the coupled Bose-Fermi mixture. The damping $\delta_k$ of the Beliaev excitations is shown as a red dash-dotted line.}  
\label{fig:densresp}
\end{figure}

Figure~\ref{fig:densresp} depicts the real part 
of the calculated density-density response for $k_{\rm F}a=-1$. 
At low frequency/momenta, we clearly see a sharp  Anderson-Bogoliubov mode where $\rm{Re}\chi(q,\omega)$ changes sign. 
The dispersion of this mode is close to the weak-coupling  result $c_{\rm s}=v_{\rm F}/\sqrt{3}$. 
For higher momenta, the dispersion curves downwards when it approaches the quasiparticle continuum starting at energies above $2\Delta$, with $\Delta \simeq 0.21\,\epsilon_{\rm F}$.

Once $\chi(q,\omega)$ is calculated, we use Beliaev theory~\cite{Fetter} to describe the effects of the resulting
 induced Bose-Bose interaction on the excitation spectrum of the atomic BEC. The single particle propagator $\bar G({\bf k},\omega)$ for the BEC
 is a $2\times2$ matrix, and the  Dyson equation reads
\begin{equation}
    \bar G({\bf k},\omega) = \bar G_0({\bf k},\omega) + \bar G_0({\bf k},\omega) \bar \Sigma({\bf k},\omega) \bar G({\bf k},\omega). 
\end{equation}
The bare propagator is
\begin{equation}
  \bar G_0({\bf k},\omega) = \left[\begin{array}{cc}
                  G_0({\bf k},\omega) & 0 \\
                  0 & G_0({\bf k},-\omega) 
              \end{array}\right],
\end{equation}
and the self-energy is
\begin{equation}
  \bar \Sigma({\bf k},\omega) = \left[\begin{array}{cc}
                  \Sigma_{11}({\bf k},\omega) & \Sigma_{12}({\bf k},\omega) \\
                  \Sigma_{21}({\bf k},\omega) & \Sigma_{11}({\bf k},-\omega) 
              \end{array}\right],
\end{equation}
where we have used  the inversion symmetry  ${\mathbf k}\leftrightarrow-{\mathbf k}$.  
The effects of interactions are included via the "Hartree-Fock" self-energies illustrated in Fig.\ \ref{UefFig} (b), and given by  
$\Sigma_{11}({\bf k},\omega) = \Sigma_{11}({\bf k},-\omega)^* = n_0 V(0,0) + n_0 V({\mathbf k},\omega)$
and
$\Sigma_{12}({\bf k},\omega) = \Sigma_{21}({\bf k},\omega) = n_0 V({\mathbf k},\omega)$.
Solving these equations for $\bar G({\bf k},\omega)$ yields the Green's functions for the diagonal elements
\begin{equation}
   G({\bf k},\omega)=\frac{\omega +\epsilon_k +n_0V({\mathbf k},\omega)}{\omega^2 - E(k,\omega)^2}
\end{equation}
where $E(k,\omega) = \epsilon_k^2 + 2\epsilon_k n_{\rm B} V(k,\omega)$.
The off-diagonal elements are $G_{12}({\bf k},\omega) =G_{21}({\bf k},\omega)= -n_{\rm B} V({\bf k},\omega)/\left[\omega^2 - E(k,\omega)^2\right]$, where $n_{\rm B}$ is the density of the BEC.
The theory satisfies the Hugenholtz-Pines relation for the chemical potential $\mu =\Sigma_{11}(0)-\Sigma_{12}(0) = n_{\rm B}V(0,0)$.
These interacting Green's functions describe excitations with energy dispersion $E_k$ given by solving 
\begin{equation}
  E_k = \mathrm{Re}\,E(k,\omega=E_k).
\label{eq:Ek}
\end{equation}
In the absence of the induced interaction, this results in the usual Bogoliubov dispersion
$\varepsilon_k = \sqrt{\epsilon_k^2 + 2n_{\rm B}{\mathcal T}_{\rm B} \epsilon_k}$.
However, due to the momentum and frequency dependence of  $V({\mathbf k},\omega)$,~\eqref{eq:Ek} is implicit and needs to be solved numerically.
The equation also yields damping of the  excitations given by $\delta_k = \mathrm{Im}\, E(k,E_k)$.

\begin{figure}
\includegraphics[width=\columnwidth]{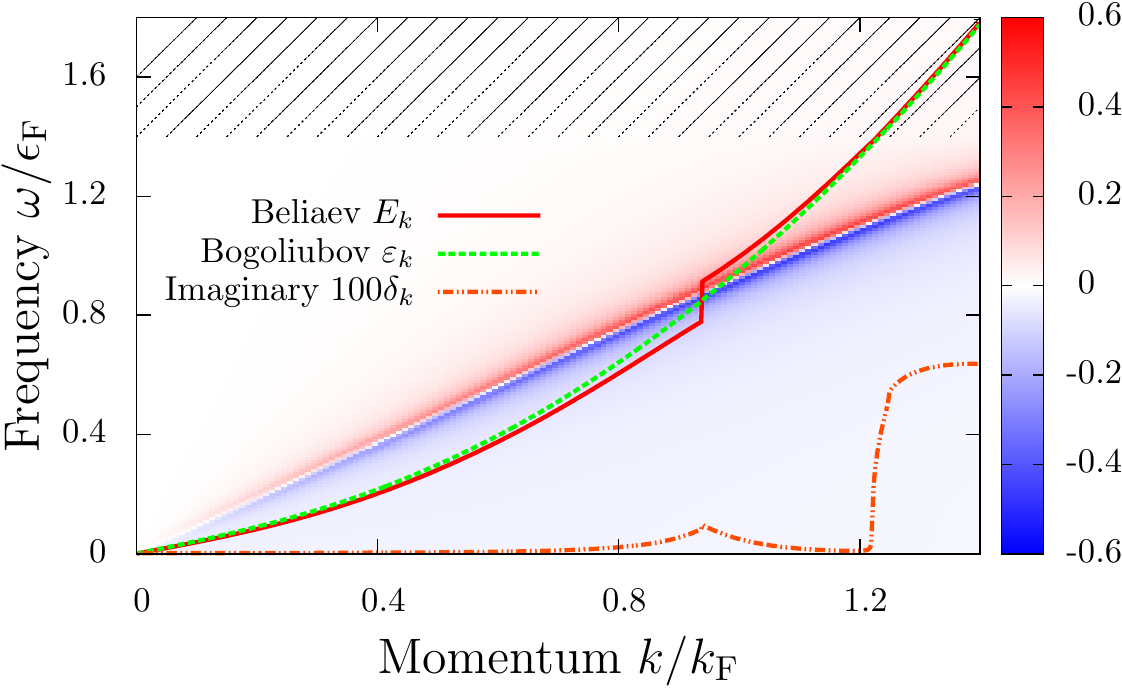}
\caption{(Color online) Similar plot as Fig.~\ref{fig:densresp} but for a unitary Fermi gas with $k_{\rm F}a_{\rm F}=\infty$. Here $\Delta \approx 0.69\,E_{\rm F}$.}
\label{fig:dispersion}
\end{figure}

\begin{figure}
\includegraphics[width=\columnwidth]{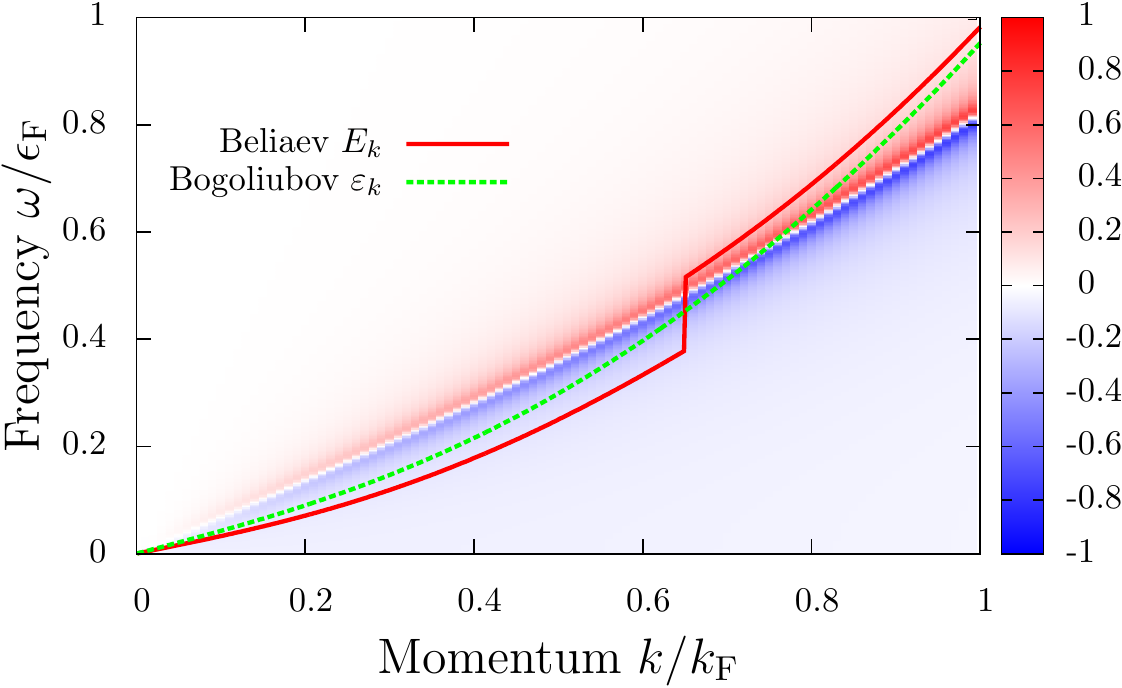}
\caption{(Color online) As Figs.~\ref{fig:densresp} and~\ref{fig:dispersion} but for the Fermi superfluid in the BEC regime with $k_{\rm F}a_{\rm F}=1$. Here $\Delta \approx 1.35\,\epsilon_{\rm F}$.}
\label{fig:dispersion_bec}
\end{figure}

Figures~\ref{fig:densresp} - \ref{fig:dispersion_bec} show the  dispersion $E_k $ obtained from~\eqref{eq:Ek}, in the BCS ($k_{\rm F}a_{\rm F}=-1$), unitarity $1/k_{\rm F}a_{\rm F}=0$, and BEC  ($k_{\rm F}a_{\rm F}=1$) regimes of the Fermi gas respectively.
The calculations are performed using parameters corresponding to: densities $n_{\rm F}=n_{\rm B} = 10^{13}\,{\rm cm}^{-3}$, scattering lengths $a_{\rm B}= a_{\rm BF} =400\,a_0$, and 
  inspired by the   superfluid Bose-Fermi mixture experiment~\cite{Barbut} we use the  masses of $^6$Li and $^7$Li  atoms. 
From~\eqref{Couplingstrength}, this yields $\kappa=4\pi a_{\rm ind}{m_{\rm B}}^{-1}v_{\rm F}^2/c_{\rm s}^2$ for the strength of the induced interaction with the effective scattering length $a_{\rm ind} \simeq 70\,a_0$.

Consider first the BCS regime with $k_{\rm F}a_{\rm F}=-1$ shown in Fig.\ \ref{fig:densresp}. 
Comparing the Bogoliubov spectrum $\epsilon_k$ for the atomic BEC decoupled from the Fermi gas with the Beliaev spectrum $E_k$ for the coupled Bose-Fermi 
mixture obtained from~\eqref{eq:Ek}, we see that coupling to the Anderson-Bogoliubov mode results in an avoided crossing. Since we are neglecting backaction effects on
 the Fermi gas, this avoided crossing becomes a discontinuous jump in the bosonic excitation frequency. We expect this prediction to be qualitatively correct, except very close 
 to the avoided crossing, since the induced interaction diverges when the two excitation frequencies are equal,   making the corresponding avoided crossing  
 sharp.   Figure \ref{fig:densresp} also shows  that the excitations of the BEC become damped when their energy is inside the quasiparticle continuum of the Fermi gas.
This reflects that the excitation dissipates energy by exciting quasiparticles in the superfluid Fermi gas. 

Figure~\ref{fig:dispersion} depicts the  spectrum $E_{k}$ when the Fermi gas is in the unitarity regime with $1/k_{\rm F}a_{\rm F}=0$. 
We again see that there is an avoided crossing, evidenced by a jump in the Beliaev dispersion $E_k$, when the Bogoliubov mode approaches the collective mode of the Fermi gas. 
In fact, the resulting energy shift is  larger than in the BCS case, since the spectral weight of the collective mode is larger in the unitarity regime.  
The bosonic excitations are again damped for energies $\omega>2\Delta$. 
The small residual damping near the avoided crossing  reflects however the small imaginary part $i\eta$ that we have built into the Fermi theory to obtain convergence.  
In the limit $\eta \rightarrow 0$, the bosonic excitations are undamped outside the quasiparticle continuum, even at the avoided crossing 
since it corresponds to the coupling of two undamped excitations.

Finally, Fig.~\ref{fig:dispersion_bec} shows the dispersion $E_k$ in the BEC regime of the Fermi gas with $k_{\rm F}a_{\rm F}=1$. 
The avoided crossing feature and the energy shift in $E_k$ is now even more pronounced due to a smaller 
sound velocity of the Fermi gas, which approaches the Bogoliubov sound speed of a dimer BEC, thereby making $\kappa$ larger as can be seen 
from~\eqref{Couplingstrength}. 
The quasiparticle continuum of the Fermi gas is outside the range of the plot due to the large pairing energy in the BEC regime. 
There is therefore no damping of the bosonic modes shown.

\begin{figure}
\includegraphics[width=\columnwidth]{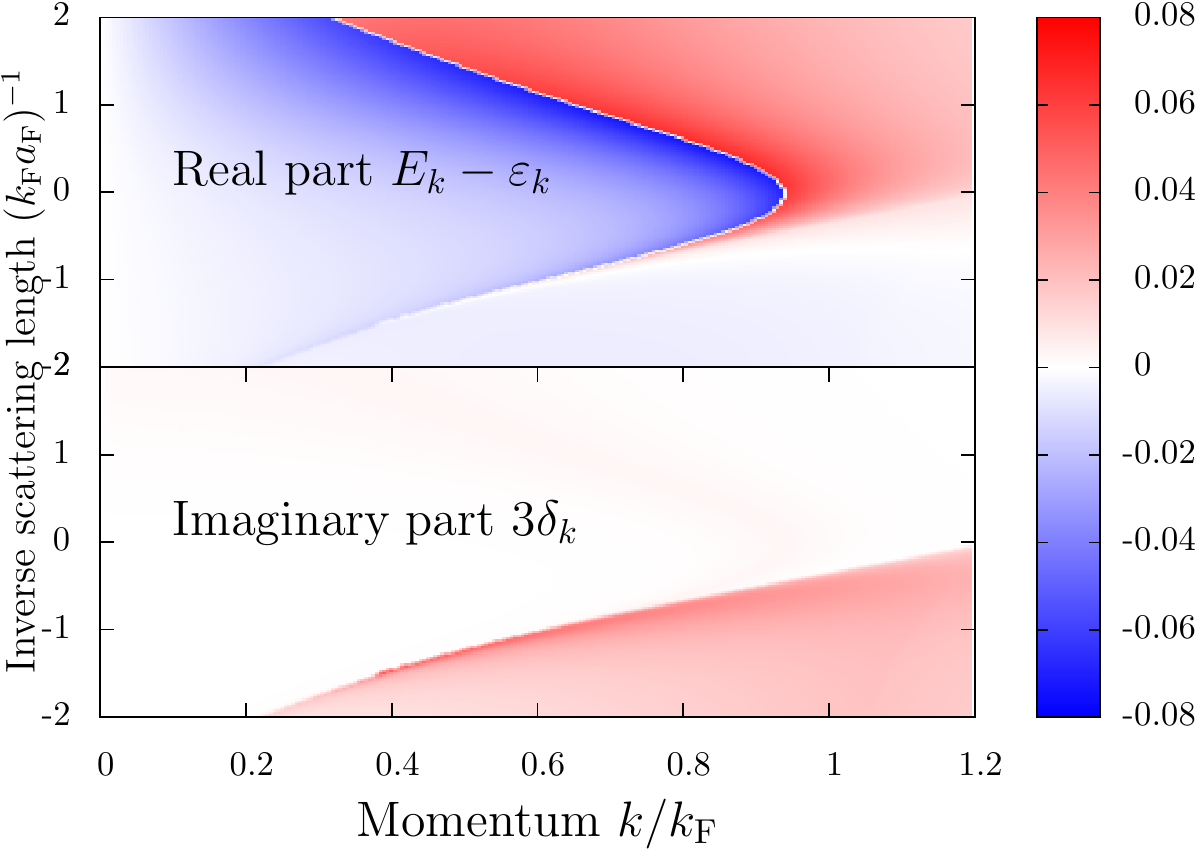}
\caption{(Color online) Top: correction $(E_k-\varepsilon_k)/\epsilon_{\rm F}$ to the BEC  dispersion  due to the induced 
interaction as a function of Fermi-Fermi scattering length $k_{\rm F}a_{\rm F}$ and momentum $k$. Bottom: decay $\delta_k/\epsilon_{\rm F}$ of the excitations in the BEC.}
\label{fig:affsweep}
\end{figure}

The above results show how the coupling between the superfluid bosons and fermions leads to significant effects on the spectrum of the atomic BEC, which  
depend on the properties of  the Fermi gas. 
In the recent experiment on the superfluid $^6$Li and $^7$Li mixture, 
the Fermi-Fermi scattering length $a_{\rm F}$ could be tuned using a Feshbach resonance. We therefore plot in Fig.~\ref{fig:affsweep} the difference $E_k - \varepsilon_k$  
between the Beliaev and Bogoliubov excitation spectra as a function of $a_{\rm F}$, keeping all other parameters as in Figs.\ \ref{fig:densresp}-\ref{fig:dispersion_bec}. 
We also plot the damping of the mode. Two effects are apparent. 
First, since the sound velocity in the Fermi gas depends on $a_{\rm F}$, the momentum where the bosonic mode exhibits the avoided crossing depends on $a_{\rm F}$.
Also, the induced interaction in general decreases/increases $E_k$ for energies below/above the avoided crossing as expected. 
The magnitude of the energy shift increases towards the BEC regime since spectral weight of the collective mode in the Fermi superfluid increases. 
Second, Fig.~\ref{fig:affsweep}  clearly shows the damping caused by the coupling to the quasiparticle excitations of the superfluid Fermi gas. 
This quasiparticle continuum moves to higher momenta as the system approaches the BEC limit and the pairing gap increases. The residual damping below the quasiparticle 
continuum shown in Fig.~\ref{fig:affsweep} is, as explained above, a result of using a non-zero $\eta$ in the numerics, and it vanished for $\eta \rightarrow 0$.
This illustrates how  the collective and single particle spectrum of the strongly correlated Fermi gas can be mapped out by measuring its effects on the  
excitations in the BEC. We note that the effects can be increased significantly by increasing $a_{\rm BF}$, since $\kappa\propto a_{\rm BF}^2$. 
In addition to varying $a_{\rm F}$ and  $a_{\rm BF}$, one can also vary $a_{\rm B}$ which will increase even further the ways one can probe the 
excitations in this Bose-Fermi mixture. The excitations of a BEC have already been measured using Bragg 
spectroscopy~\cite{Kozuma1999,Stenger1999,StamperKurn1999,Steinhauer2002,Ozeri2005,Papp2008,Kinnunen2009,Ronen2009}. 

In conclusion, we examined a mixture of a BEC and a superfluid Fermi gas using Beliaev theory for the bosons combined with quasiparticle random phase approximation for the fermions. 
The fermions were shown to mediate a frequency/momentum dependent interaction between the bosons, which leads to two qualitatively new effects. 
First, the induced interaction diverges at the sound mode of the Fermi gas which results in a sharp avoided crossing feature  in the excitation spectrum of the BEC. 
Second, the excitation of quasiparticles in the Fermi gas leads to a damping of the excitations of the BEC. 
By varying the densities and scattering lengths of the system, these effects can be used to systematically probe the properties of the Fermi gas in the strongly correlated BCS-BEC cross-over. 
Our work may be extended in a number of directions: It would be interesting to include  
the backaction of the bosons on the superfluid Fermi gas to obtain a detailed description of the avoided crossing of the sound modes. 
Trapping effects can be included using a local density approximation, which has proven to work well when considering short wavelength Bragg scattering~\cite{StamperKurn1999}. Finally, the theory can be extended to finite temperatures, which would result in a damping of BEC excitations for all momenta due to the presence of thermally excited quasiparticles in the Fermi gas.

\begin{acknowledgments}
JJK acknowledges support by the Academy of Finland through its Centres of Excellence Programme (2012-2017) under Project No. 251748. 
GMB would like to acknowledge the support of the Hartmann Foundation via grant A21352 and the Villum Foundation via grant VKR023163.
\end{acknowledgments}

\end{document}